# The factor paradox: Common factors can be correlated with the variance not accounted for by the common factors!

André Beauducel[1]

**Abstract**

The case that the factor model does not account for all the covariances of the observed variables is considered. This is a quite realistic condition because some model error as well as some sampling error should usually occur with empirical data. It is shown that principal components representing covariances not accounted for by the factors of the model can have a non-zero correlation with the common factors of the factor model. Non-zero correlations of components representing variance not accounted for by the factor model with common factors were also found in a simulation study. Based on these results it should be concluded that common factors can be correlated with variance components representing model error as well as sampling error. In consequence, even when researchers decide not to represent some small or trivial variance by means of a common factor, these excluded variances can still be part of the model.

---

[1] Institute of Psychology, University of Bonn, Kaiser-Karl-Ring 9, 53111 Bonn, Germany, Email: beauducel@uni-bonn.de

## 1 Introduction

The factor model has been developed primarily in psychology and has meanwhile been applied to a broad variety of data in many fields, even outside of psychology. A merit of the model is that latent variables can be constructed that may explain the covariances between observed variables. The model has been described in several books (e.g., Gorsuch, 1983; Harris, 2001; Harman, 1976; Tabachnick & Fidell, 2007) and several software packages are available for the estimation of the model parameters. Nevertheless, it might not be regarded as a realistic assumption that the factor model fits perfectly to the data (MacCallum, 2003; MacCallum & Tucker, 1991). MacCallum and Tucker (1991) called the misfit of the factor model in the population *model error* and they also describe sources of sampling error for the factor model. Both model error and sampling error might have the effect that the factor model does not account for the complete covariance of observed variables. Whereas MacCallum (2003) as well as MacCallum and Tucker (1991) were concerned with the consequences of model error and sampling error for the model description and for the estimation of model parameters, the present paper investigates the correlation of the variance not accounted for by the factor model with the common factors. Beauducel (2013) found that the correlation between the variance not accounted for by the principal component model and the common factors is not necessarily zero. This means that variance that is regarded as irrelevant according to principal component analysis can be relevant for the common factors. In contrast, the present paper investigates the correlation of the variance not accounted for by the factor model with the common factors. Since this variance is not part of the factor model one would expect this variance to be uncorrelated with the common factors.

## 2 Definitions

The defining equation of the common factor model is

$$\mathbf{x} = \mathbf{\Lambda}\mathbf{f} + \mathbf{e}, \qquad (1)$$

where **x** is the random vector of observations of order $p$, **f** is the random vector of factor scores of order $q$, **e** are the unobservable random error vectors or error factor scores of order $p$, and $\Lambda$ is the factor pattern matrix of order $p$ by $q$. The common factor scores **f**, and the error factor scores **e** are assumed to have an expectation zero ($\varepsilon(\mathbf{x}) = 0$, $\varepsilon(\mathbf{f}) = 0$, $\varepsilon(\mathbf{e}) = 0$). The expected variance of the factor scores is one, the covariance between the common factors and the error factors is assumed to be zero ($\text{Cov}(\mathbf{f}, \mathbf{e}) = \varepsilon(\mathbf{fe}') = 0$). The expected covariance matrix of observed variables $\Sigma$ can be decomposed into

$$\Sigma = \Lambda \Phi \Lambda' + \Psi^2, \tag{2}$$

where $\Phi$ represents the $q$ by $q$ factor correlation matrix and $\Psi^2$ is a $p$ by $p$ diagonal matrix representing the expected covariance of the error factors **e** ($\text{Cov}(\mathbf{e}, \mathbf{e}) = \varepsilon(\mathbf{ee}') = \Psi^2$). Moreover, postmultiplication of Equation 1 with **e**´ shows that the expected covariance of the error factors with the observed variables is $\text{Cov}(\mathbf{e}, \mathbf{x}) = \varepsilon(\mathbf{ex}') = \Psi^2$, because $\varepsilon(\mathbf{fe}') = \mathbf{0}$. It is assumed that the diagonal of $\Psi^2$ contains only positive values so that $\Psi^2$ is nonsingular.

## 3 Results

### 3.1 Correlation with unexplained variance

Consider that there is some model error so that the factor model in the population does not account for the covariance of the observed variables completely. This could be written as

$$\Sigma = \Lambda \Phi \Lambda' + \Psi^2 + \Omega, \tag{3}$$

with $\Omega$ representing the expectation of the residual covariances. Since the factor model as defined in Equation (2) accounts for all diagonal elements in $\Sigma$, there are only nondiagonal elements in $\Omega$. The remaining definitions of the factor model are not altered by Equation (3), only the nonzero nondiagonal elements in $\Omega$ are taken into account. Let $\Omega = \mathbf{KVK}'$ with $\mathbf{KK}' = \mathbf{K}'\mathbf{K} = \mathbf{I}$ be the eigen-decomposition of $\Omega$, where **V** is diagonal with the eigenvalues in decreasing order. Since the main-diagonal of $\Omega$ contains only zero values, the trace of **V** will be zero. Therefore, even when some nondiagonal elements of $\Omega$ are not zero, only the first eigenvalues of $\Omega$ will be positive and some negative eigenvalues will also occur. In the following, only the eigenvectors corresponding to positive eigenvalues are considered. The matrix $\mathbf{K}^*$ contains only eigenvectors corresponding to positive eigenvalues and $\mathbf{V}^*$ contains only positive eigenvalues (in descending order), so that $\mathbf{N} = \mathbf{K}^*\mathbf{V}^{*1/2}$. $\mathbf{N}$ is called the loading matrix of the corresponding principal components. $\mathbf{\Omega}^*$ is the matrix of expected residual covariances reproduced from the principal components with positive eigenvalues, with

$$\mathbf{\Omega}^* = \mathbf{NN}´ = \mathbf{K}^*\mathbf{V}^*\mathbf{K}^{*}´. \tag{4}$$

Accordingly, the corresponding residuals of the observed variables can be decomposed into principal components $\mathbf{u}$, which yields

$$\mathbf{x} - \mathbf{\Lambda}\mathbf{f} - \mathbf{e} = \mathbf{Nu}, \tag{5}$$

with $\mathbf{u}$ being orthogonal components with $\varepsilon(\mathbf{uu}´) = \mathbf{I}$. It follows from Equation (5) and from the definitions of the factor model that

$$\varepsilon(\mathbf{xf}´) = \varepsilon(\mathbf{\Lambda\Phi} + \mathbf{Nuf}´) \tag{6}$$

and that

$$\varepsilon(\mathbf{xe}´) = \varepsilon(\mathbf{\Psi}^2 + \mathbf{Nue}´). \tag{7}$$

The following theorem states that when the expected correlation between the error factor scores $\mathbf{e}$ and the residual components $\mathbf{u}$ is zero, a nonzero expected correlation of the common factors $\mathbf{f}$ with $\mathbf{u}$ occurs.

**Theorem 3.1.** $\varepsilon(\mathbf{fu}´) \neq \mathbf{0}$ *if* $\varepsilon(\mathbf{eu}´) = \mathbf{0}$.

**Proof.** It follows from Equations (5), (6), and (7) that

$$\mathbf{NN}´ = \varepsilon(\mathbf{\Sigma} - (\mathbf{\Lambda\Phi} + \mathbf{Nuf}´)\mathbf{\Lambda}´ - \mathbf{\Psi}^2 - \mathbf{Nue}´ - \mathbf{\Lambda}(\mathbf{\Lambda\Phi} + \mathbf{Nuf}´)´ + \mathbf{\Lambda\Phi\Lambda}´ - \mathbf{\Psi}^2 - \mathbf{eu}´\mathbf{N}´ + \mathbf{\Psi}^2)$$

$$= \varepsilon(\mathbf{\Sigma} - \mathbf{\Lambda\Phi\Lambda}´ - \mathbf{Nuf}´\mathbf{\Lambda}´ - \mathbf{\Psi}^2 - \mathbf{Nue}´ - \mathbf{\Lambda\Phi\Lambda}´ - \mathbf{\Lambda fu}´\mathbf{N}´ + \mathbf{\Lambda\Phi\Lambda}´ - \mathbf{eu}´\mathbf{N}´)$$

$$= \varepsilon( - \mathbf{Nuf}´\mathbf{\Lambda}´ - \mathbf{Nue}´ - \mathbf{\Lambda fu}´\mathbf{N}´ - \mathbf{eu}´\mathbf{N}´). \tag{8}$$

For $\varepsilon(\mathbf{eu}´) = \mathbf{0}$ Equation (8) can be transformed into

$$\mathbf{NN}´ = \varepsilon( - \mathbf{Nuf}´\mathbf{\Lambda}´ - \mathbf{\Lambda fu}´\mathbf{N}´ ). \tag{9}$$

Equation (9) is true *iff* $\varepsilon(\mathbf{\Lambda fu}´) = -0.5\ \mathbf{N}$. This completes the proof. $\square$

Theorem 3.2 specifies the condition for the expected zero correlation of the common factors $\mathbf{f}$ with the residual components $\mathbf{u}$.

**Theorem 3.2.** $\varepsilon(\mathbf{fu}') = \mathbf{0}$ *iff* $\varepsilon(\mathbf{eu}') = -0.5\ \mathbf{N}$.

**Proof.** For $\varepsilon(\mathbf{fu}') = \mathbf{0}$ Equation (8) can be transformed into

$$\mathbf{NN}' = \varepsilon(-\mathbf{Nue}' - \mathbf{eu}'\mathbf{N}'). \tag{10}$$

Equation (10) is true *iff* $\varepsilon(\mathbf{eu}') = -0.5\ \mathbf{N}$. This completes the proof. □

Theorem 3.3 specifies that the expected correlation of the variance accounted for by the factor model $(\Lambda\mathbf{f} + \mathbf{e})$ with the residual components $\mathbf{u}$ is not zero.

**Theorem 3.3.** $\varepsilon((\Lambda\mathbf{f} + \mathbf{e})\mathbf{u}') \neq \mathbf{0}$.

**Proof.** Equation (8) can be transformed into

$$\mathbf{NN}' = \varepsilon(-\mathbf{Nu}(\Lambda\mathbf{f} + \mathbf{e})' - (\Lambda\mathbf{f} + \mathbf{e})\mathbf{u}'\mathbf{N}'). \tag{11}$$

Equation (11) is true *iff* $\varepsilon((\Lambda\mathbf{f} + \mathbf{e})\mathbf{u}') = -0.5\ \mathbf{N}$. This completes the proof. □

The meaning of Theorem 3.3 is that the components $\mathbf{u}$, representing variance not accounted for by the factor model, have a nonzero correlation with the variance accounted for by the factor model.

Transformation of Equation 5 reveals that the components $\mathbf{u}$ can be calculated as

$$\mathbf{u} = (\mathbf{N}'\mathbf{N})^{-1}\mathbf{N}'(\mathbf{x} - \Lambda\mathbf{f} - \mathbf{e}). \tag{12}$$

Both $\mathbf{f}$ and $\mathbf{e}$ are usually unknown due to factor score indeterminacy (Guttman, 1955) so that Equation 12 implies that indeterminacy also holds for $\mathbf{u}$. Thus, in a typical situation of an applied researcher, the correlation of the components representing the variance not accounted for by the factor model with the common factors remains unknown.

### 3.2 Simulation Study

The magnitude of the correlations between the common factors and the components representing the residuals cannot be investigated within empirical studies, because the population common factor scores and the population error factor scores are unknown. However, in simulation studies population common factor scores and population error factor scores can be fixed a priori so that their correlation with the components representing residuals can be investigated in the population and in the sample. Therefore, a simulation study was conducted in order to investigate the size of the correlations between the common factors and components representing the variance that is not accounted for by the factor model. It is, moreover possible to investigate in the simulation study the case that the residual covariances do not represent model error but sampling error. This is interesting because sampling error should not be systematically related to the common factors.

Only the first component representing the variance not accounted for by the factor model was considered in this simulation, because the first component will always summarize most of the residual variance and because it will always have a positive eigenvalue. Of course, in most data sets even more than one component with eigenvalues greater than zero will occur. However, it is already informative to investigate the correlation of the first component of residuals with the common factors. The conditions of the small simulation study were the number of cases in the sample (150, 300, 900 cases), the size of the salient loadings (.40, .60, .80), and the number of factors (3, 6 factors). For each of the 18 conditions (3 numbers of cases x 3 loading sizes x 2 number of factors) 1,000 factor analyses were performed with SPSS 18. The population for the models comprised 900,000 cases. As an example the population three-factor model for salient loadings of .40 is presented in Table 1.

Table 1: Three-factor population models based salient loadings of .40

| variables | orthogonal | | |
|---|---|---|---|
| | F1 | F2 | F3 |
| $x_1$ | .40 | .00 | .00 |
| $x_2$ | .40 | .00 | .00 |
| $x_3$ | .40 | .00 | .00 |
| $x_4$ | .40 | .00 | .00 |
| $x_5$ | .40 | .00 | .00 |
| $x_6$ | .00 | .40 | .00 |
| $x_7$ | .00 | .40 | .00 |
| $x_8$ | .00 | .40 | .00 |
| $x_9$ | .00 | .40 | .00 |
| $x_{10}$ | .00 | .40 | .00 |
| $x_{11}$ | .00 | .00 | .40 |
| $x_{12}$ | .00 | .00 | .40 |
| $x_{13}$ | .00 | .00 | .40 |
| $x_{14}$ | .00 | .00 | .40 |
| $x_{15}$ | .00 | .00 | .40 |

Since orthogonal population models were investigated, Varimax-rotation (Kaiser, 1958) was performed for the 18,000 maximum likelihood factor analyses based on the random samples drawn from the population. The factor analyses were based on the correlations of the observed variables.

The mean eigenvalues of the first components representing the correlations not accounted for by the factor model were larger for smaller sample sizes and for smaller salient loading sizes (see Table 2). This was to be expected, since no model error was present in the population models so that the first component should only represent residual correlations that are due to sampling error.

Table 2: Means and standard deviations (in brackets) of eigenvalues of the first principal component calculated from the residual correlations

| Salient loading | Sample size | 3 factors M | (SD) | 6 factors M | (SD) |
|---|---|---|---|---|---|
| .40 | 150 | .66 | (.67) | 2.02 | (1.02) |
|     | 300 | .17 | (.13) | .54 | (.51) |
|     | 900 | .05 | (.01) | .11 | (.02) |
| .60 | 150 | .16 | (.04) | .36 | (.07) |
|     | 300 | .08 | (.02) | .17 | (.03) |
|     | 900 | .02 | (.01) | .05 | (.01) |
| .80 | 150 | .08 | (.03) | .20 | (.04) |
|     | 300 | .04 | (.01) | .10 | (.02) |
|     | 900 | .01 | (.00) | .03 | (.01) |

The distributions of the correlations of the first component representing the residual correlations of factor analysis with the first common factor are presented in Figure 1. Although the distributions are quite symmetric, the kurtosis of the distribution of correlations was a bit smaller when based on the six-factor solutions (Figure 1 A) than the kurtosis of the distribution of correlations when based on the three-factor solutions (Figure 1 B).

(A)
frequency

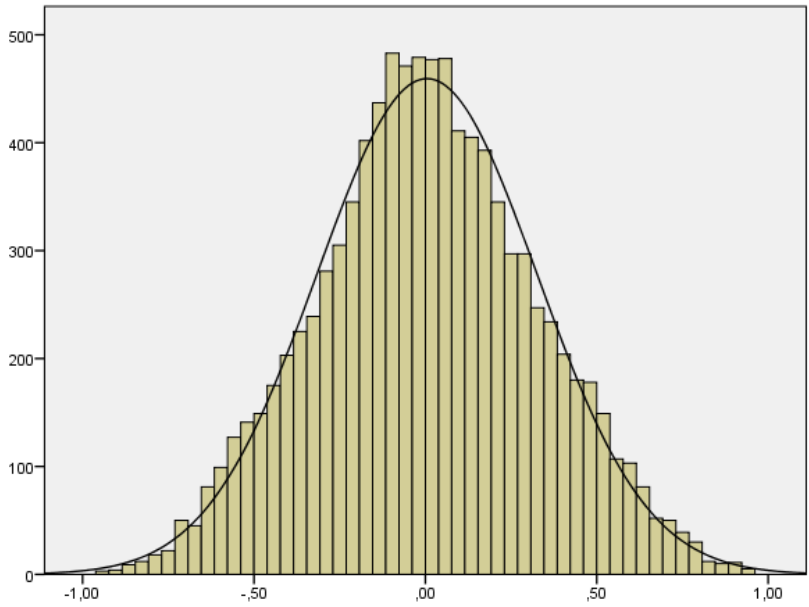

r

(B)
frequency

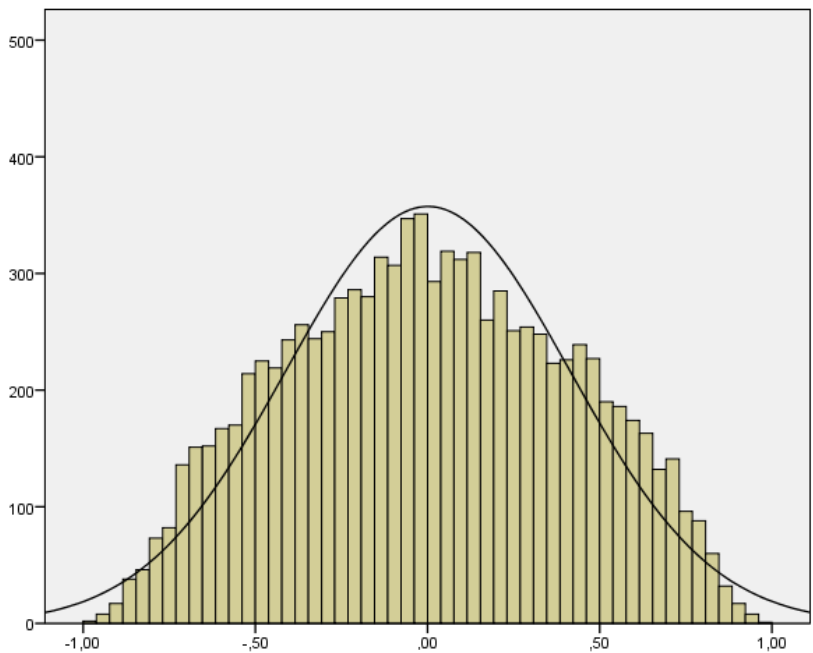

r

Figure 1. Histogram of correlations between the first component of residuals with the first common factor in the analysis based on three factors (A) and the analysis based on six factors (B).

More importantly, the whole range of positive and negative correlations occurred: 17,1% of the correlations had an absolute size greater than .80 for the three-factor solutions and 2,9% of the correlations had an absolute size greater than .80 for the six-factor solutions. In order to provide a more complete description of the effects of the conditions of the simulation study, the root mean squared correlation (RMSC) of the first component representing the residual correlations from factor analysis with the common factors was computed. No results of significance tests were reported because –due to the large sample size of the simulation study– all condition main effects and all interactions were significant in ANOVA at the .001 level. The size of the relevant effects can be depicted from Figure 2: The RMSC was larger than .50 for the three factor solutions and was about .40 for the six factor solutions. Thus, the RMSC decreases with the number of factors, but it does not decrease with sample size and with the size of the salient loadings.

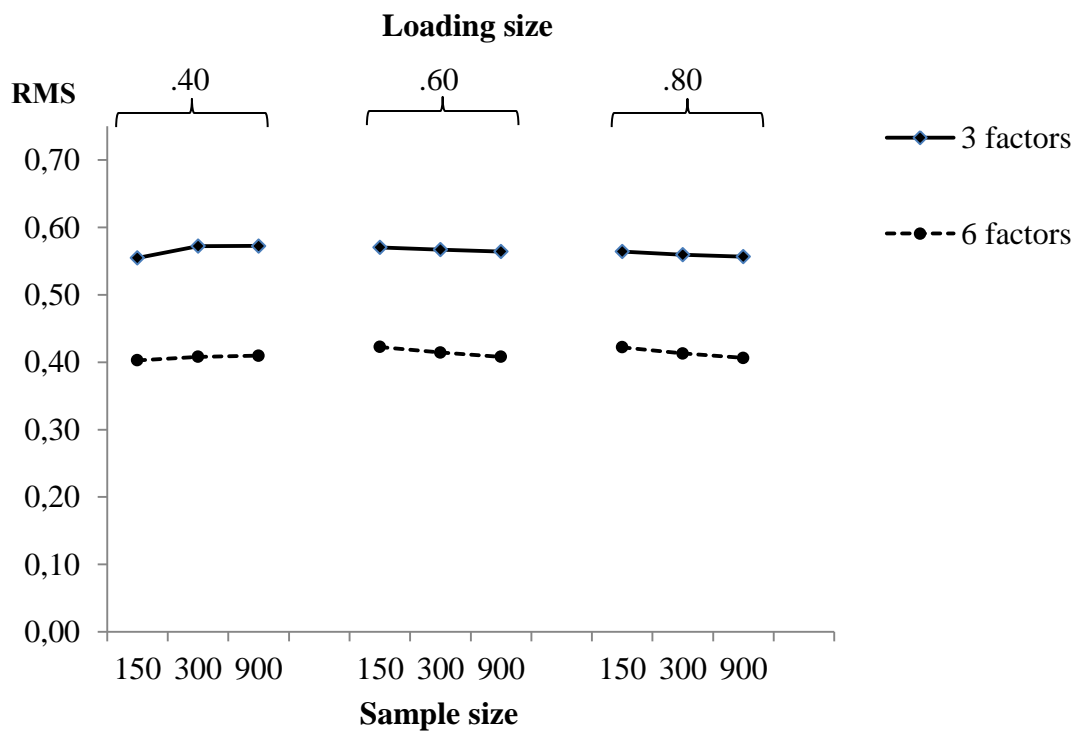

Figure 2. Root mean squared correlation (RMS) of the first component representing residuals of factor analysis with common factors for the conditions of the simulation. Standard errors were smaller than .01 and were therefore not presented.

## Discussion

The correlation of the variance not accounted for by the factor model with the common factors was investigated. Since this unexplained variance is by definition not part of the factor model one would expect components representing this variance to be uncorrelated with the common factors. However, it was shown algebraically and by means of a simulation study that the common factors can have in fact a nonzero correlation with components representing the variance not accounted for by the factor model. Moreover, the sum of the common factor variance and the error factor variance representing the total variance that is accounted for by the factor model was shown to have a nonzero correlation with the variance not accounted for by the factor model. According to Theorem 3 the common factors are uncorrelated with the components representing unexplained variance only under a condition implying that the error factors have a nonzero correlation with these components. However, since the error factors are aimed at representing unique variances, they should not be correlated with any other variance according to the factor model.

Moreover, a simulation study revealed that the root mean squared correlation of the first component representing unexplained variance with the common factors was greater than .50 for the three-factor solutions and about .40 for the six-factor solutions. It should be concluded that these correlations cannot be regarded as being virtually zero in general. Since factor analysis will nearly always be performed on sample data and since this would always lead to some covariances that are not accounted for by the factor model (MacCallum, 2003; MacCallum & Tucker, 1991), the results presented here will be relevant for most applications of factor analysis. Overall, it seems that further methodological developments are necessary in order to provide a form of factor analysis that avoids the problem that has been demonstrated here. Meanwhile, severe caution is recommended with respect to the interpretation of common factors as representing only the common variance of the observed variables. In contrast, it is likely that the common factors are correlated with the variance that is not accounted for by the factors.